\documentclass[conference]{IEEEtran}
\IEEEoverridecommandlockouts
\usepackage[style=ieee]{biblatex}
\usepackage{amsmath,amssymb,amsfonts}
\usepackage{algorithmic}
\usepackage{graphicx}
\usepackage{textcomp}
\usepackage{xcolor}
\def\BibTeX{{\rm B\kern-.05em{\sc i\kern-.025em b}\kern-.08em
    T\kern-.1667em\lower.7ex\hbox{E}\kern-.125emX}}
\begin{document}

\title{Fully Convolutional Network for Removing DCT Artefacts From Images
\thanks{This work was supported by the Polish National Science Centre under grant no. 2017/27/B/ST6/02852 and the project financed under the program of the Polish Minister of Science and Higher Education under the name ``Regional Initiative of Excellence'' in the years 2019--2022 project number 020/RID/2018/19, the amount of financing 12,000,000.00 PLN.}
}

\author{\IEEEauthorblockN{Patryk Najgebauer, Rafa{\l} Scherer}
\IEEEauthorblockA{
\textit{Cz\c{e}stochowa University of Technology}\\
Cz\c{e}stochowa, Poland \\
\{patryk.najgebauer, rafal.scherer\}@pcz.pl 
}
\and
\IEEEauthorblockN{Leszek Rutkowski, Fellow, IEEE}
\IEEEauthorblockA{
\textit{Information Technology Institute} \\
\textit{University of Social Sciences}\\
90-113 Lodz, Poland
}
}

\maketitle

\begin{abstract}
Image compression is one of the essential methods of image processing. Its most prominent advantage is the significant reduction of image size allowing for more efficient storage and transfer. However, lossy compression is associated with the loss of some image details in favor of reducing its size. In compressed images, the deficiencies are manifested by noticeable defects in the form of artifacts; the most common are block artifacts, ringing effect, or blur.
In this article, we propose three models of fully convolutional networks with different configurations and examine their abilities in reducing compression artifacts. In the experiments, we research the extent to which the results are improved for models that will process the image in a similar way to the compression algorithm, and whether the initialization with predefined filters would allow for better image reconstruction than developed solely during learning.
\end{abstract}

\begin{IEEEkeywords}
denoising, compresion artefact, JPEG, FCN, DCD
\end{IEEEkeywords}

\section{Introduction}
Currently, image lossy compression algorithms are essential, and their significance is much higher than before. Since the creation of social networking, a tremendous amount of information describing daily events has been generated and presented in the form of multimedia content. These extensive data are not overwritten or deleted over time, and more are added each day. This forces the continuous expansion of the data storage and processing infrastructure, and without compression, the infrastructure had to be many times larger, which involves high investment costs. Another measurable effect of compression is a significant reduction in network bandwidth, which is also important nowadays.

Generally, there are two forms of image compression: lossy and lossless. In the lossless one like PNG (Portable Network Graphics), we usually deal with computer graphics that do not contain too much detail, and we care about the perfect reproduction of the original image. 
Currently, the most commonly used are real-world images in the form of photographs or movies. These images are characterized by a high level of detail so that lossless compression is no longer as effective. However, a large part of these details, especially small ones, is only the background of the context, and people do not pay attention to them. Even if they are removed from the image, the image is still readable.

We expect compression algorithms to be accurate and fast. Despite the development of new image compression algorithms, the block JPEG algorithm based on the DCT (Discrete Cosine Transform) is still the most commonly used. Its advantage is simplicity and widespread use. It is used as a part of many video \cite{rafajlowicz2010testing} compression algorithms and had an implementation of hardware support in many devices.

The use of lossy image compression involves choosing the compression ratio that makes an impact on the final image quality. Most often, the choice is due to hardware restrictions in the form of available space for storing photos or the bandwidth of the transportation medium. Usually, in high-resolution digital photographs, increasing the degree of compression does not significantly affect the quality of the image. It is different in the case of images with a low resolution where the effect of increasing the compression level is manifested by the formation of visible artifacts and distortions.

Along with the development of machine learning methods, more and more effective methods are created that allow to denoise the image and improve its quality. These techniques may include methods for reducing the grain resulting from the underexposure of the camera sensor. Another solution is methods of image deblurring \cite{nah2017deep, tao2018scale} to remove the effect of camera shake while taking the picture.
Also interesting are methods of superresolution \cite{yamanaka2017fast, dong2016accelerating} that increase the image resolution with context prediction of enlarged details.
Notable in this field are also methods allowing to restore the original image after lossy compression \cite{dong2015compression, svoboda2016compression}, which main task is to remove compression artifacts.
Together with the super-pixel methods, they allow improving the quality of digital photographs taken from years ago as well as to enlarge the selected part of the image and present it in a well-looking form.

Artificial neural networks from the very beginning were used to denoise the image \cite{zhou1988image}, then as deep neural networks \cite{xie2012image}. Currently, the best capabilities in the field of image analysis are provided by convolutional neural networks \cite{burger2012image}. In most cases, the standard CNN model output is composed of fully connected layers that mix and analyze the whole context of the image \cite{Duda2019Training}. Networks of this type are perfectly suitable for classification purposes. For more local image context processing such as object instance classification \cite {girshick2015fast, ren2015faster, rafajlowicz2008local,ke2019neuro}, segmentation \cite {he2017mask, pinheiro2015learning} better results are obtained by networks without fully-connected layers, i.e. fully convolutional networks (FCN) \cite {long2015fully}. In FCNs, a combination of convolutional and deconvolution layers \cite {noh2015learning} gives excellent results in image local context analysis and presenting it in the form of spatial representation, for example for image reconstruction \cite{wei2019regional, wei2012combined}.

\begin{figure}
\centering
\includegraphics[width=0.9\columnwidth]{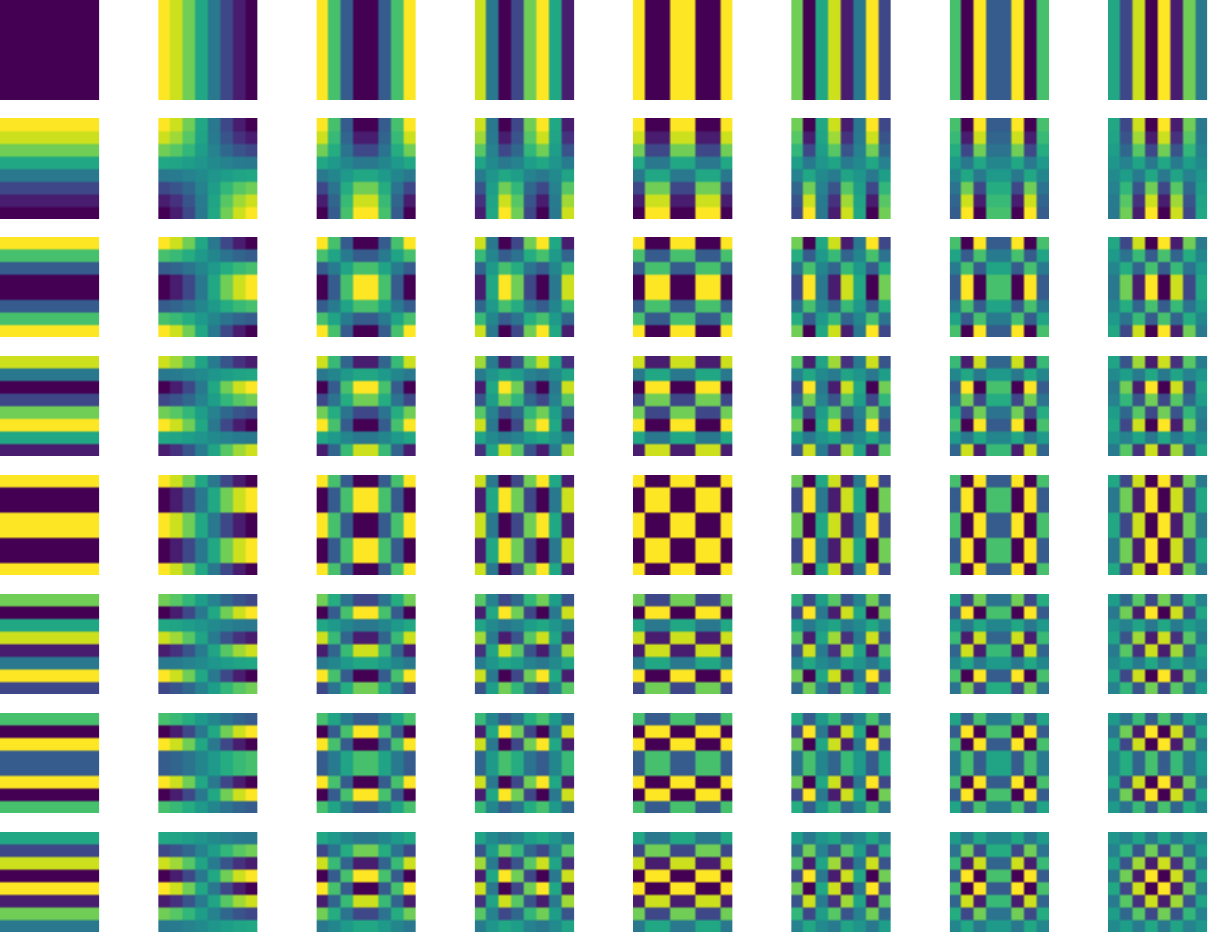}
\caption{DCT patterns visualization. The combination of the presented 64 patterns with their coefficients allows reproducing any image without loss.}
\label{fig:dct_filters}
\end{figure}

%
As mentioned earlier, lossy image compression can significantly reduce the size of the resulting image, however, at the cost of losing the ability to restore it to its original form. In the paper, we focused on the JPEG algorithm. The general principle of this algorithm is to divide the image into $8 \times 8$ pixel blocks that are independently processed separately for each channel. Each block is sampled with a set of discrete cosine transform (DCT) convolutions (Fig. \ref{fig:dct_filters}) and then represented by the determined coefficients. In this form, there is no loss of image details, and it can be reproduced in the reverse operation by assembling the image from the DCT filters and their coefficients. For the JPEG algorithm, the compression ratio is regulated by the quality $q$, which refers to the accuracy of the DCT coefficients. A higher compression ratio results in less accuracy of preserved coefficients.

\begin{figure}
\centering
\includegraphics[width=\columnwidth]{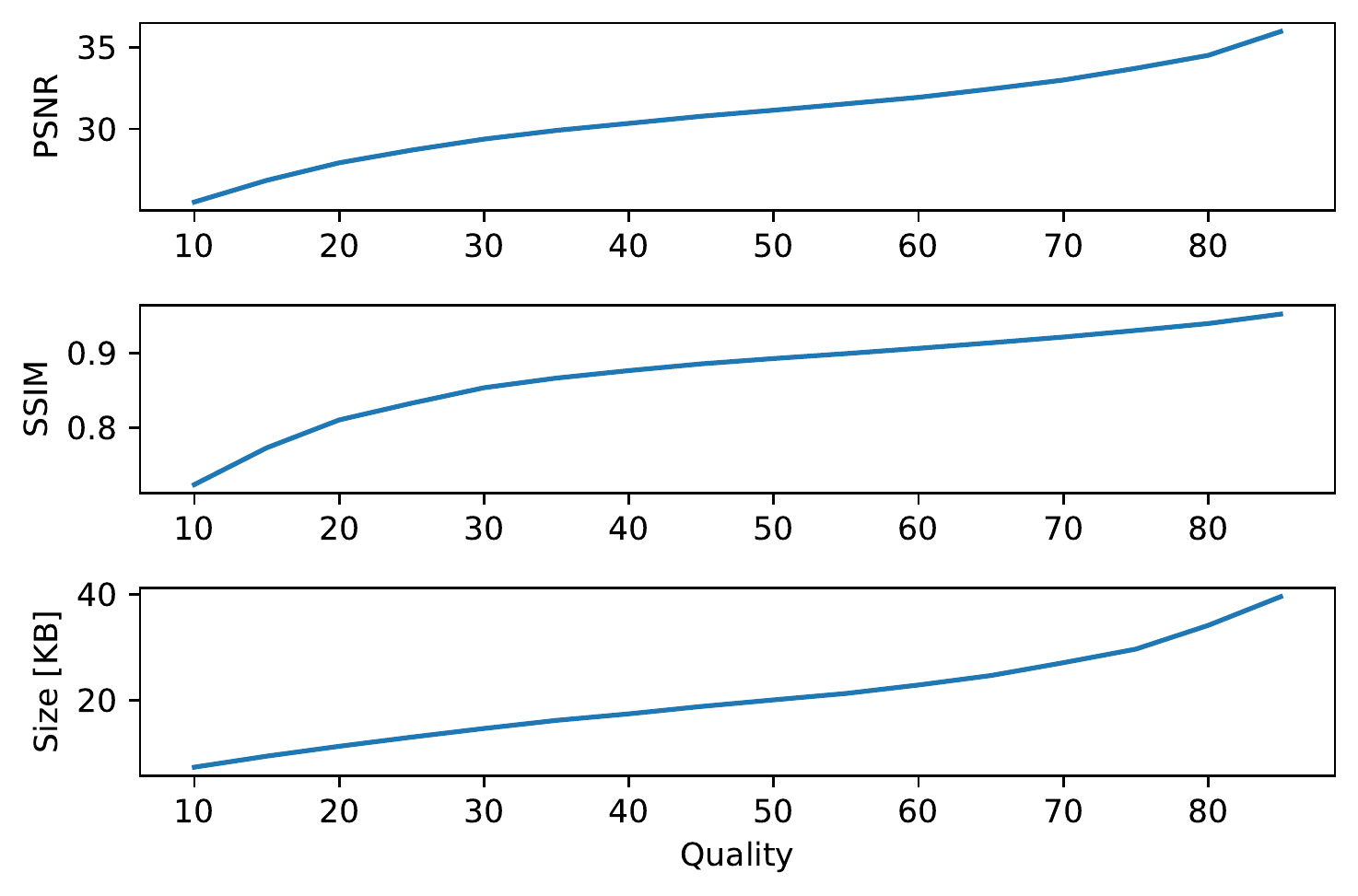}
\caption{Example graphs of the relationship between the quality of the JPEG compression and the size of the compressed file and the PSNR and SSIM measures related to the original image.}
\label{fig:encode_pnsr_ssim}
\end{figure}

During decompression, not accurate coefficient values applied to filters bring some deviate between the restored image and its origin in the form of visible artifacts. For any image operations, especially compression and decompression, the most common measures of quality are the peak signal-to-noise ratio (PSNR) and the structural similarity index measure (SSIM) \cite{hore2010image}. These measures objectively evaluate image divergence after conversion relatively to its original.  Measures are based on a comparison of two images. In many cases, the effectiveness of these measures depends on the accuracy of the reference image. It is especially important in the image improvement where the reference image may be more corrupt and noisy than the effect obtained.  
Figure \ref{fig:encode_pnsr_ssim} shows a graph of file size changes corresponding to the peak signal-to-noise ratio (PSNR) and the structural similarity index measure (SSIM) depending on the value of the quality factor.
The values of image quality measures and the file size change along with the compression ratio.

\begin{figure*}
\centering
\includegraphics[width=\textwidth]{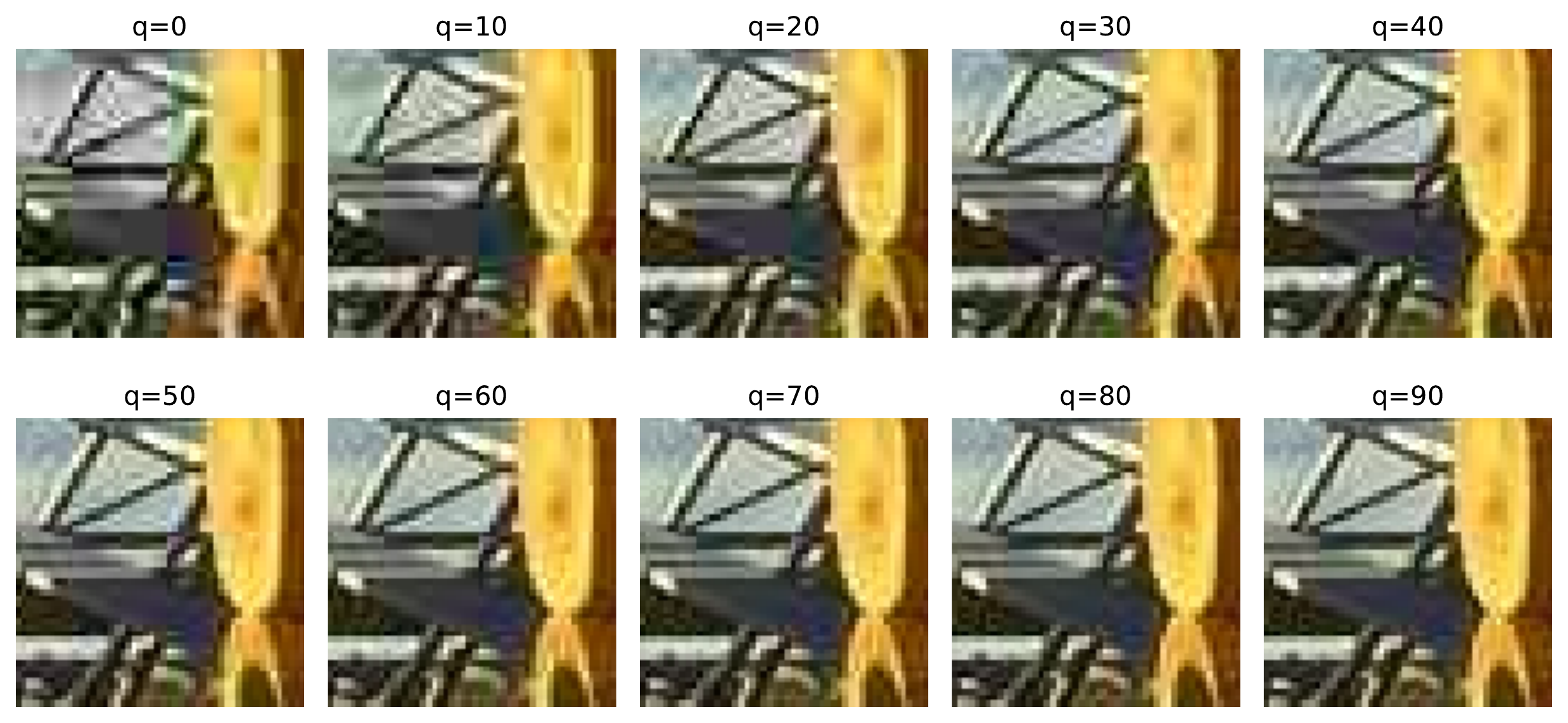}
\caption{Example of an image with various compression level. The example presents an increase in artifact visibility with changing compression quality $q$.}
\label{fig:encode_example}
\end{figure*}

The disparities of image decompression are on the entire surface of the image; however, it should be noted that most of them are not visible to the human eye and do not interfere with the reading of the image context. Human perception does not pay much attention to areas of the image with a high frequency of changes, such as small leaves on trees; these areas are generalized. However, human vision is more focused on the edges of highly contrasting areas. In these places, all compression imperfections are very noticeable, as presented in Figure \ref{fig:encode_example}. 
At the edges of these areas, the variance of the DCT coefficient increases significantly, which makes the artifacts even more visible. In contrast, the variance of the coefficients over uniform and homogeneous areas is small, making the artifacts practically invisible.

In this paper, we propose new FCN models for DCT-based compressed image improvement. Moreover, we examine whether the convolutional layers with predefined, non-learned coefficients will improve the results of removing artifacts related to JPEG lossy compression. The proposed models are relatively small and accurate. 
Through this research, we highlight the following features and contributions of the proposed models.
\begin{itemize}
\item We present the first neural networks with DCT layers. 
\item Our work provides new insights, showing that neural networks with DCT layers can be smaller and more accurate than plain fully convolutional networks. 
\item Predefined DCT coefficients can be more universal in JPEG improvements than all-trained image filters. 
\end{itemize}
The remainder of the paper is organised as follows.
 In Section \ref{sec:method}, we described the proposed neural network architectures. Section \ref{sec:exper} describes experiments that show the accuracy and comparison of the models and the ARCNN model.
 Finally, conclusions and discussions of the paper are presented in Section \ref{sec:conclusions}.

\section{Novel FCN Models for Compressed Image Improvement}
\label{sec:method}
As aforementioned, we propose new FCN models for compressed image improvement. Moreover, we examine whether the convolutional layers with predefined, non-learned coefficients will improve the results of removing artifacts related to JPEG lossy compression.
For this purpose, we created and trained three network models and, as a reference, compared them with the AR-CNN \cite{dong2015compression} network. 
In the case of networks described in the literature, a huge emphasis was placed on the size of the network to be relatively small and fast. 
The AR-CNN network, in comparison to CAS-CNN \cite{cavigelli2017cas}, L08 \cite{svoboda2016compression}, SR-CNN \cite{dong2015image} obtains the best results in terms of the precision and size of the model. It is also the closest to the models presented in the article.

We must also mention the modification of the AR-CNN network, i.e. Fast AR-CNN \cite{yu2016deep}, where authors insert an additional layer of $1 \times 1$ kernel size. This significantly reduced the size of the entire network to 57,296 parameters. The second change is using two-pixels stride of the first layer convolutions sampling that accelerated the algorithm several times in comparison to the AR-CNN network
The fast version of the network obtains worse results, thus in the presented paper the comparison was made with the AR-CNN network.
All the examined models are fully convolution networks which depth does not exceed four layers. Table \ref{tab:models_config} presents a summary of the number of layers and the number of parameters of the analyzed networks.
\begin{figure}
\centering
\includegraphics[width=\columnwidth]{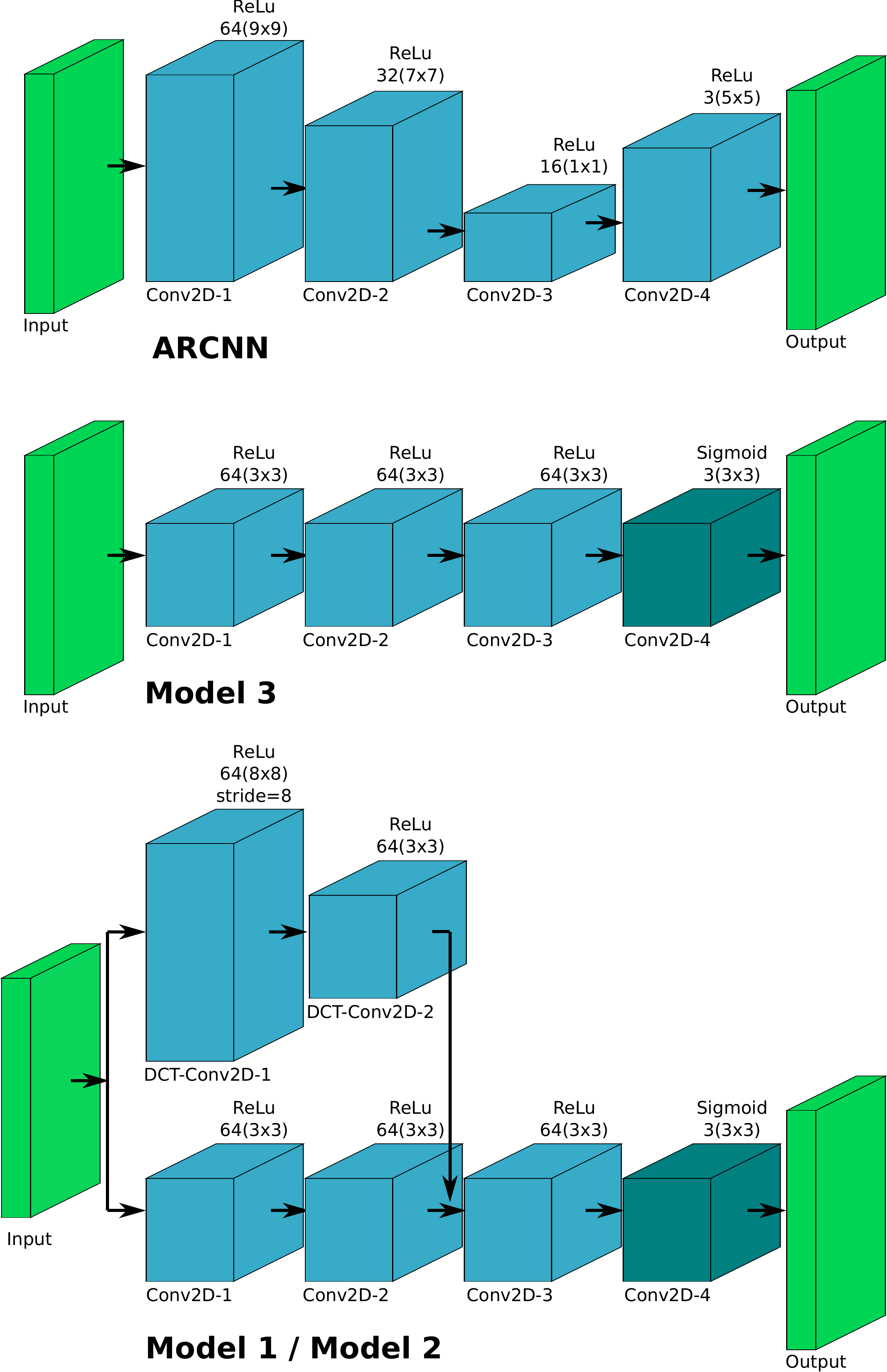}
\caption{Structures of fully convolutional network models presented in the paper. Model 1 consists of four convolution layers of $3 \times 3$ kernel size. Model 2 has an additional residual block which the first layer has a kernel size of $8 \times 8$. Model 3 is similar to Model 2, but with the first additional layer with the weights initiated by the DCT transform and not subjected to training.}
\label{fig:model}
\end{figure}
%
\begin{table}[htbp]
\caption{Meta-parameters of the proposed FCN models (Models 1-3) and the ARCNN model.}
\begin{center}
\begin{tabular}{ccrc}
\hline 
\textbf{Model} & \textbf{Layer} & \textbf{No. of }    &  \textbf{Filters/Kernel size/Stride/}\\
							&   & \textbf{params} &	 \textbf{Activation/Padding} \\
\hline
\textbf{Model 1} & DCT-Conv2D-1   & 12,288 & 64/8/8/ReLu/valid \\
 & DCT-Conv2D-2   & 36,928 & 64/3/1/ReLu/same \\
 & Conv2D-1   & 1,792 & 64/3/1/ReLu/same \\
 & Conv2D-2   & 36,928 & 64/3/1/ReLu/same \\
 & Conv2D-3   & 73,792 & 64/3/1/ReLu/same \\
 & Conv2D-4        & 1,731 & 3/3/1/Sigmoid/same \\
\textbf{total} & \textbf{6} & \textbf{163,459}  \\
\hline
\textbf{Model 2} & DCT-Conv2D-1   & 12,288 & 64/8/8/ReLu/valid \\
 & DCT-Conv2D-2   & 36,928 & 64/3/1/ReLu/same \\
 & Conv2D-1   & 1,792 & 64/3/1/ReLu/same \\
 & Conv2D-2   & 36,928 & 64/3/1/ReLu/same \\
 & Conv2D-3   & 73,792 & 64/3/1/ReLu/same \\
 & Conv2D-4   & 1,731 & 3/3/1/Sigmoid/same \\
\textbf{total} & \textbf{6} & \textbf{163,459}  \\
\hline
\textbf{Model 3} & Conv2D-1   & 1,792 & 64/3/1/ReLu/same \\
 & Conv2D-2   & 36,928 & 64/3/1/ReLu/same \\
 & Conv2D-3   & 36,928 & 64/3/1/ReLu/same \\
 & Conv2D-4   & 1,731 & 3/3/1/Sigmoid/same \\
\textbf{total} & \textbf{6} & \textbf{77,379}  \\
\hline
\textbf{ARCNN} & Conv2D-1   & 5,184 & 64/9/1/ReLu/same \\
 & Conv2D-2   & 100,352 & 32/7/1/ReLu/same \\
 & Conv2D-3   & 512 & 16/1/1/ReLu/same \\
 & Conv2D-4   & 400 & 3/5/6/ReLu/same \\
\textbf{total} & \textbf{6} & \textbf{106,448}  \\
\hline
\end{tabular}
\label{tab:models_config}
\end{center}
\end{table}
Model 3 and AR-CNN are networks with four layers. The AR-CNN in the tested version has the first layer with 64 convolutions of $9 \times 9$ size, then 32 of $7 \times 7$ for data shrink, 16 of $1 \times 1$ for enhancement, and the last layer with 3 of $5 \times 5$ kernel size convolutions for reconstructing. Uncommon, wider kernel sizes allow for a small number of layers to cover a sufficient area of the image to eliminate artifacts from the image effectively.
Our Model 3 (Fig. \ref{fig:model}) consists of three identical layers of 64 $3 \times 3$ convolutions and the last of 3 of $3 \times 3$ kernel size convolutions with the sigmoid activation function. In the case of our network, a much smaller area is generalized to predict a single pixel of an image.

Model 1  and 2 (Fig. \ref{fig:model}) are similar to Model 3, but with the additional residual block. The block itself consists of two layers, the second layer of standard 64 $3 \times 3$ convolution filters, but the first DCT layer consists of 64 filters of $8 \times 8$ kernel sizes.  The first layer samples the image with 8-pixel stride, which means that it examines every 8-pixel block individually without overlap (8-pixel stride). Image sampling by this layer is identical to the DCT filters of the JPEG compression algorithm. In Model 2, the DCT layer is subjected to regular training to develop its weights.
\section{Experimental Results}
\label{sec:exper}
In the experiments, to compare our models with the literature, we used the BSDS500 dataset \cite{amfm_pami2011} containing about 500 natural photographs $481\times 321$ pixels. In lower-resolution images, compression distortions are much more noticeable than in high-resolution images. The training set was presented as the ground truth set, while its compressed version became the input set. For the experiment, the learning set was compressed with quality $q=60$. The models are trained to remove artifacts that remain after compression; therefore, only image slices are presented as the input because the wider context of the image is not needed for the proper learning in our case. The input fragments of the image also are not pre-processed or scaled that the compression blocks perfectly match the DCT layer filter. The image to be improved is processed in fragments and, finally, assembled into one the whole image.

\begin{figure}
\centering
\includegraphics[width=\columnwidth]{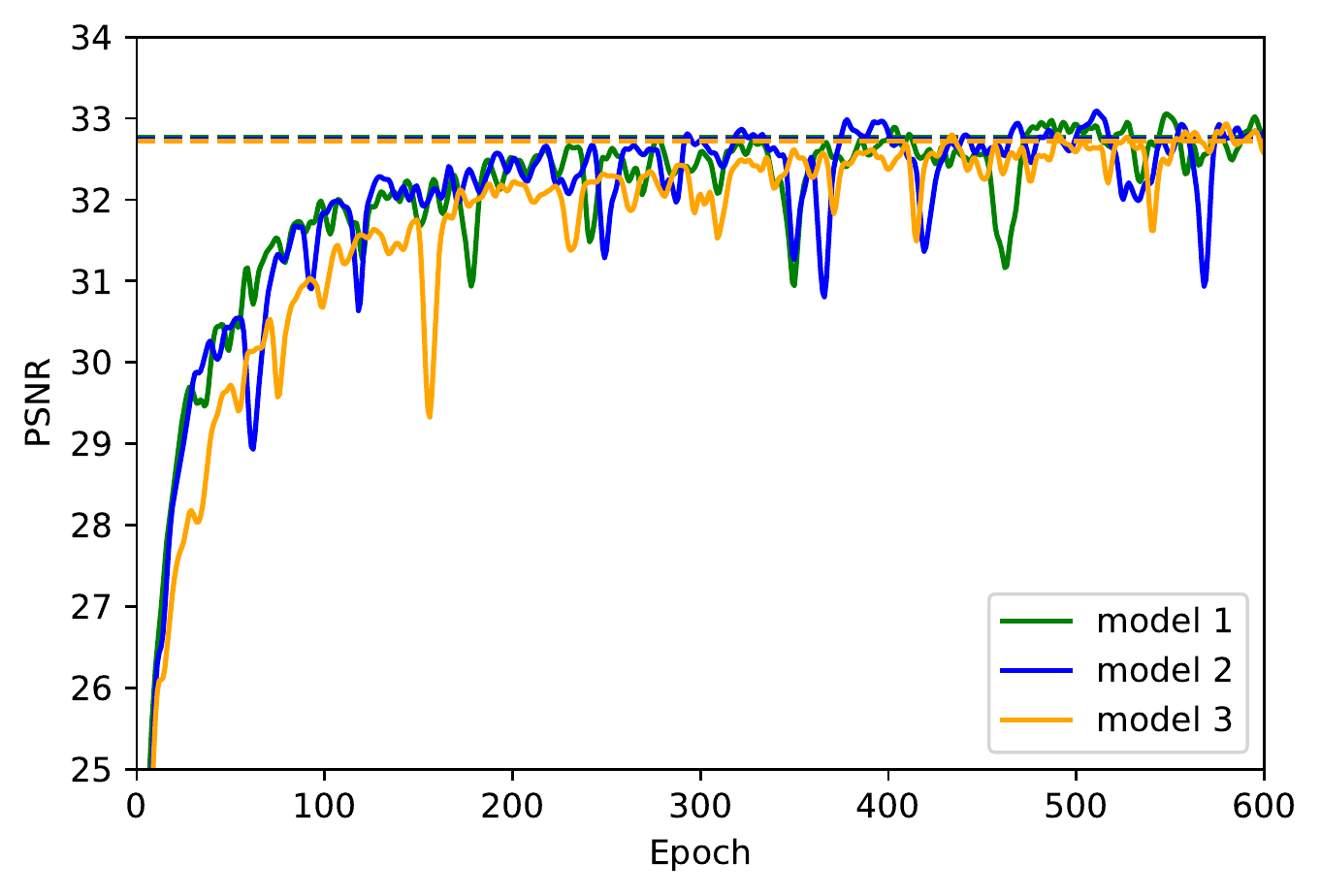}
\caption{Accuracy of the proposed model training over each training epoch with the average accuracy from the last twenty epochs.}
\label{fig:train_graph}
\end{figure}

The models were trained with images with three RGB channels. During training, the presented models obtained a similar average PSNR value (about 32.8 dB) to the original images from the learning set.
Fig. \ref{fig:train_graph} presents the training process in individual epochs. The graph fluctuates because, during learning, random sets of fragments of compressed images were characterized by varying degrees of noise compared to their uncompressed originals.

\begin{figure}
\centering
\includegraphics[width=\columnwidth]{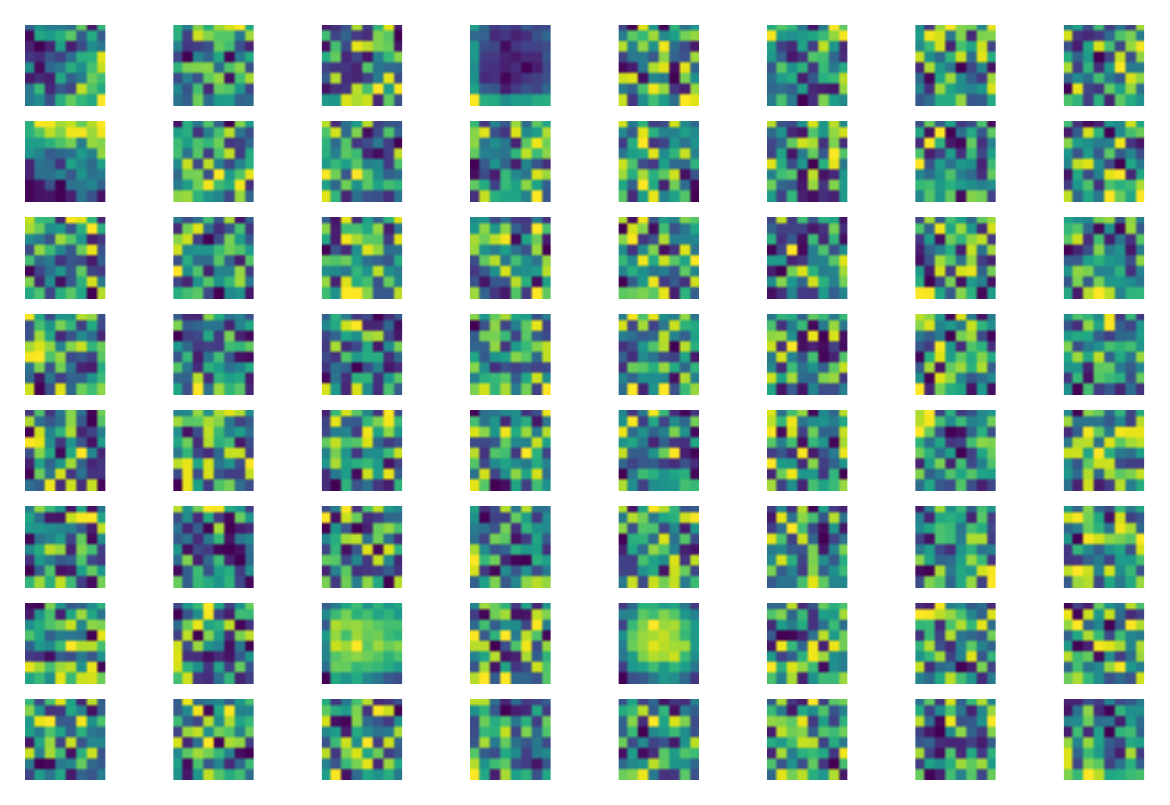}
\caption{Visualization of the trained weights of the first layer of the residual block in Model 2. The learned filters lost the universality of the DCT filters from Fig. \ref{fig:dct_filters} by accommodating to a particular domain. }
\label{fig:dct_train_filters}
\end{figure}
\begin{figure}
\centering
\includegraphics[width=\columnwidth]{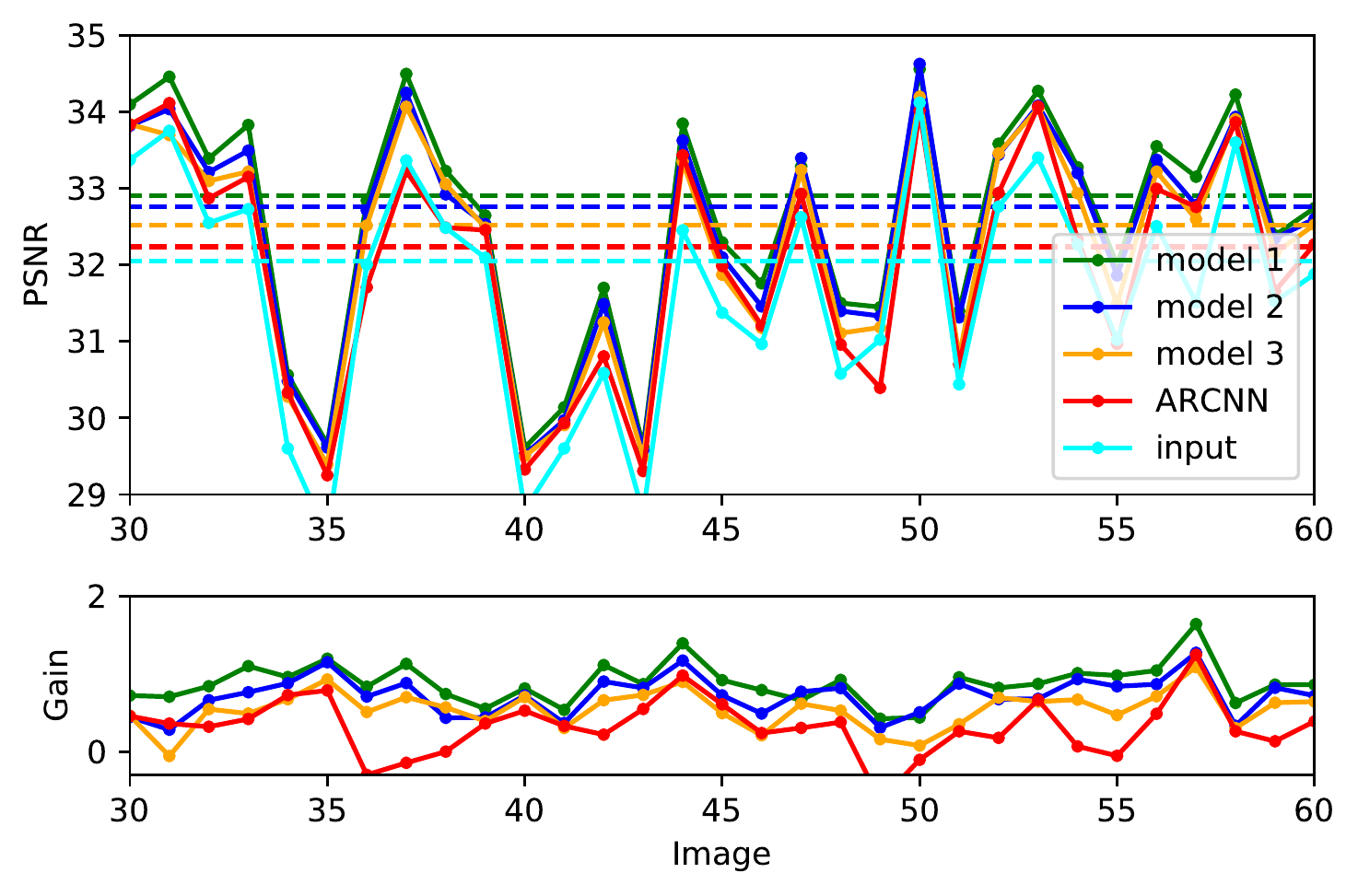}
\caption{Graph of the accuracy of the artifact reduction for individual images from the test set measured with the PSNR metric.}
\label{fig:result_psnr}
\end{figure}

\begin{figure}
\centering
\includegraphics[width=\columnwidth]{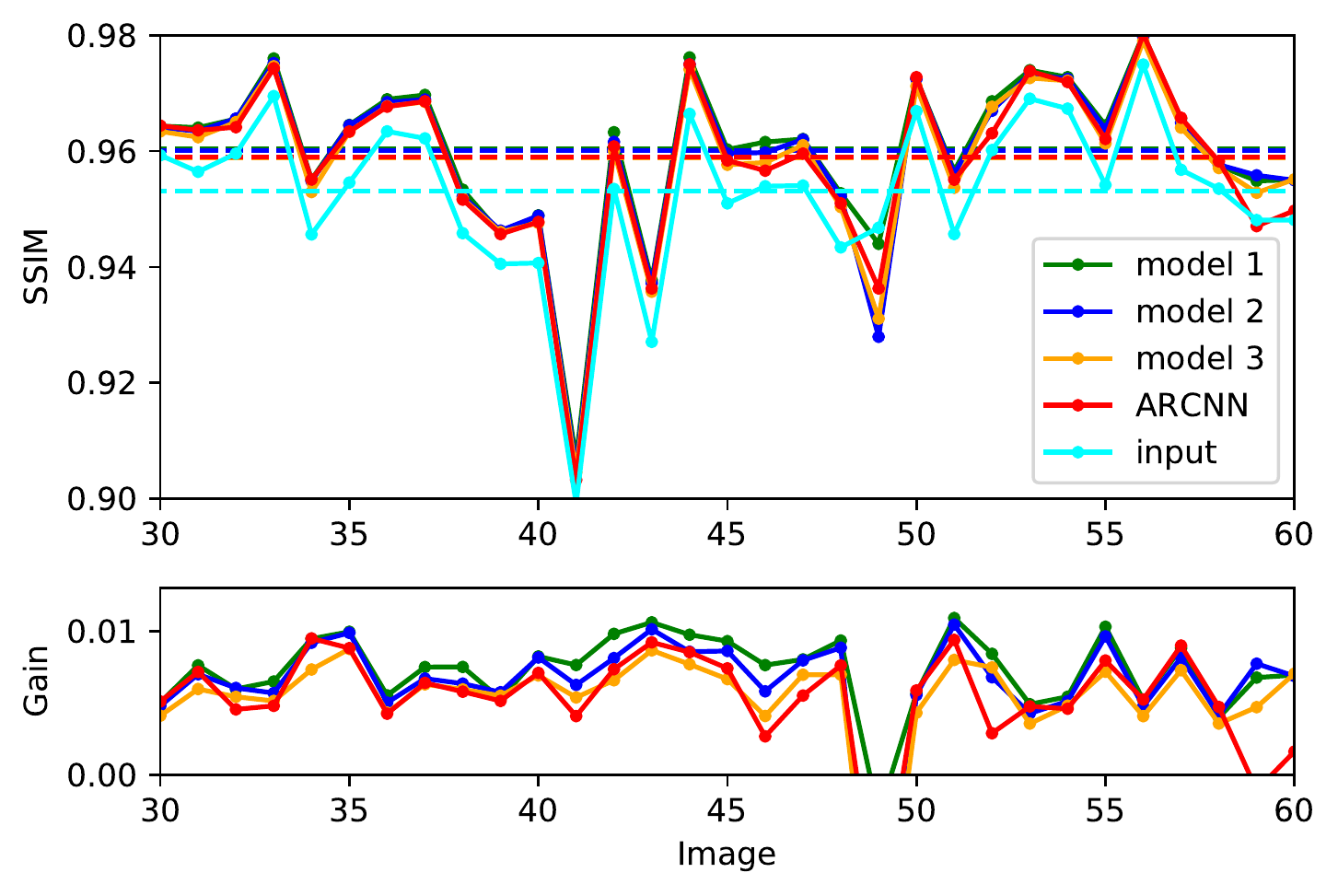}
\caption{Graph of the accuracy of artifact reduction for individual images of test set measured with the SSIM.}
\label{fig:result_ssim}
\end{figure}

\begin{table}[htbp]
\caption{Table of the accuracy of artifact reduction for the test set measured with the PSNR metric.}
\begin{center}
\begin{tabular}{ccccccc}
\hline 
\textbf{Model} & \multicolumn{3}{c}{\textbf{PSNR}} & \textbf{Worst result} & \textbf{Best result} \\
\              & min & avg & max & & \\
\hline
model 1 & 28.02 & 32.90 & 39.09 & 346016.jpg & 70011.jpg \\
model 2 & 28.04 & 32.76 & 38.06 & 346016.jpg & 43051.jpg \\
model 3 & 27.91 & 32.51 & 38.32 & 346016.jpg & 70011.jpg \\
ARCNN & 27.85 & 32.23 & 38.84 & 346016.jpg & 70011.jpg \\
input & 27.43 & 32.05 & 37.83 & 346016.jpg & 70011.jpg \\
\hline
\multicolumn{6}{c}{\textbf{Gain}} \\
\hline
model 1 & -0.26 & 0.86 & 2.11 & 196027.jpg & 48017.jpg \\
model 2 & -0.04 & 0.72 & 1.68 & 250087.jpg & 48017.jpg \\
model 3 & -1.49 & 0.47 & 1.55 & 250087.jpg & 48017.jpg \\
ARCNN & -3.05 & 0.19 & 1.36 & 250087.jpg & 48017.jpg \\
\hline
\end{tabular}
\label{tab:results_psnr}
\end{center}
\end{table}
In the case of the test set, inconsistencies are obtained between each model. These are slight differences that cannot be directly seen in the image. Block artifacts by each model have been removed and are no longer noticeable. From the results of the PSNR noise presented in Table \ref{tab:results_psnr}, it can be seen that Model 1 with predefined weights obtained the best average results, next was Model 2 where the indicated layer was learned. It can be assumed that the predefined layer (Fig. \ref{fig:dct_filters}), in this case, gets better results because it was more universal than the learned layer which visualization of weights is presented in Fig. \ref{fig:dct_train_filters}. Accuracies of the artifact reduction measured by PNSR and SSIM are presented in Fig. \ref{fig:result_psnr} and \ref{fig:result_ssim}, respectively. 

\begin{figure}[htbp]
\centering
\includegraphics[width=\columnwidth]{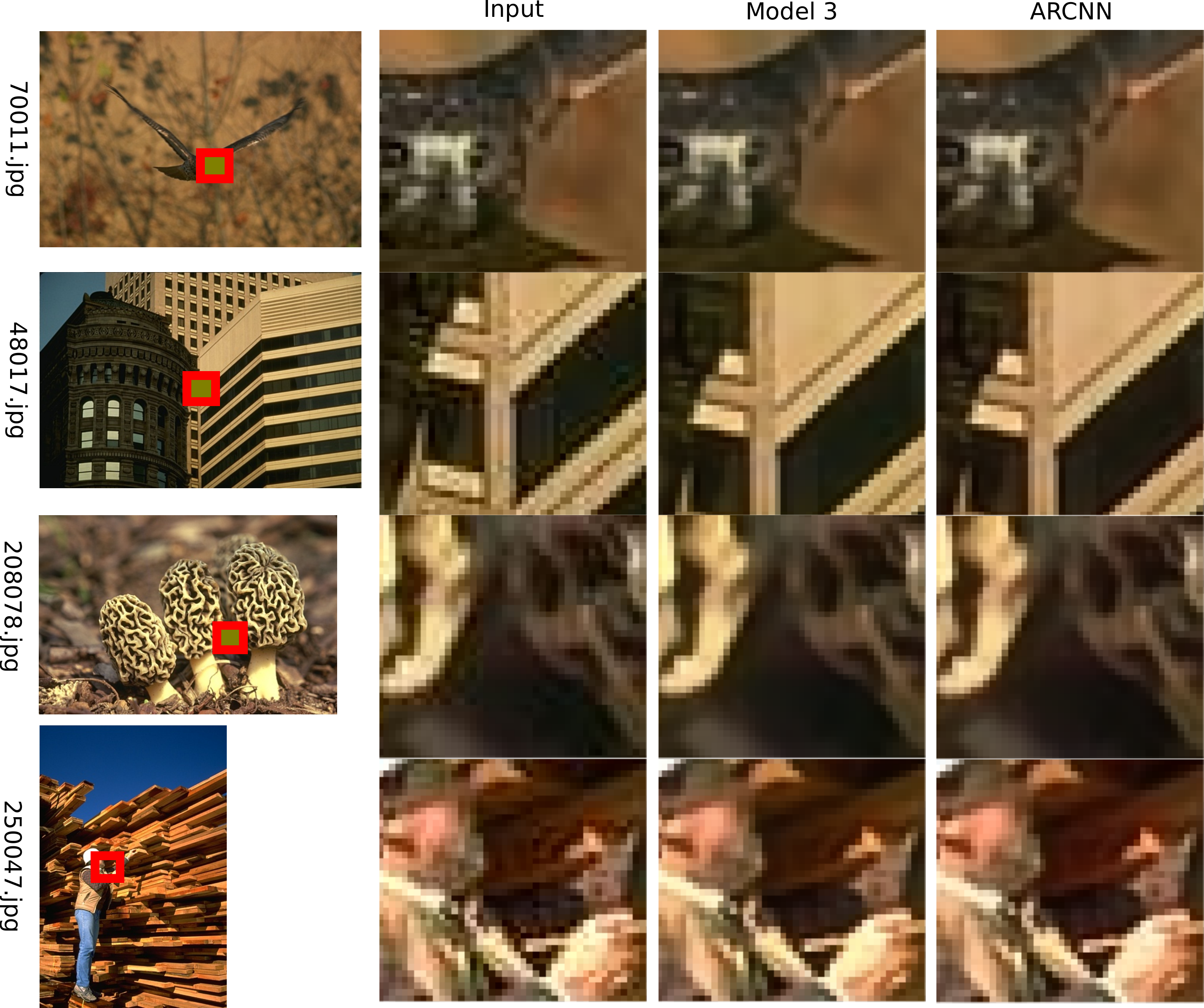}
\caption{Example best cases of artifact reduction.}
\label{fig:best_results}
\end{figure}

\begin{figure}[htbp]
\centering
\includegraphics[width=\columnwidth]{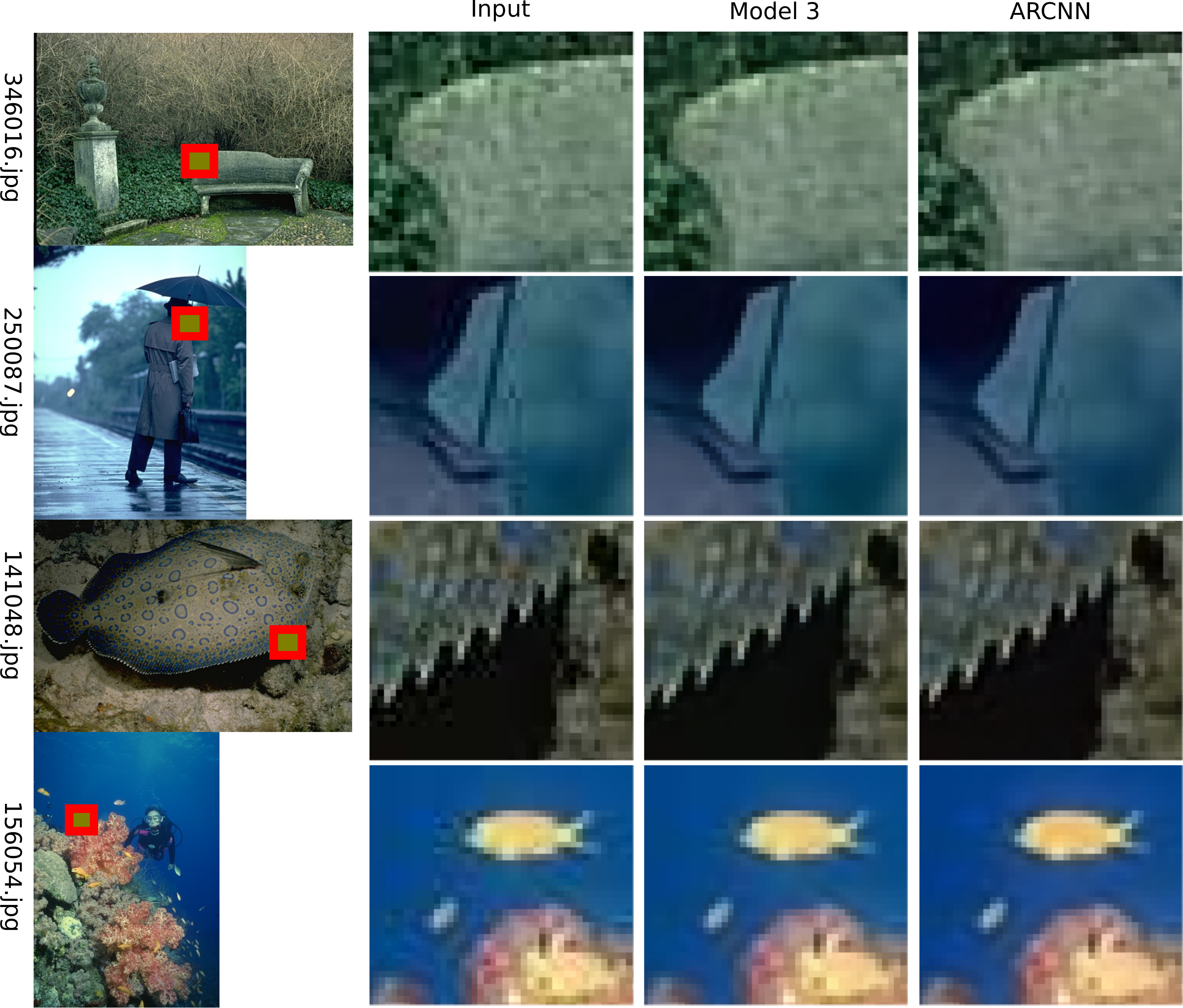}
\caption{Example worst cases of artifact reduction.}
\label{fig:worst_results}
\end{figure}

In the case of SSIM from Tab. \ref{tab:results_ssnr}, the results of all algorithms are very close to each other and practically identical, which indicates that among other block artifacts have been removed with similar efficiency.
Fig. \ref{fig:best_results} shows the images for which the best results were obtained for both PSNR and SSIM measures. These are images that contain a lot of uniform surfaces and blur caused by shallow focus (a small depth of field). There are no areas with a granular texture. After compression, the artifacts are especially visible in these images; however, they were successfully removed. 
Fig. \ref{fig:worst_results} shows the worst results obtained. For the PSNR measure, these are images that have a lot of small details like small branches or rain. These details were significantly damaged by the compression itself, and even the removal of the artifacts, their old details cannot be reproduced. In the case of SSIM, the worst results were obtained on images that also had a lot of details but not as small as in the case of PSNR. These were mainly textures that were partially blurred under compression, and its elements were often larger than a single $8\times8$ pixel field (CNN filter of DCT coefficient size).

\begin{table}[htbp]
\caption{Table of the accuracy of artifact reduction for test set measured with the SSIM.}
\begin{center}
\begin{tabular}{ccccccc}
\hline 
\textbf{Model} & \multicolumn{3}{c}{\textbf{SSIM}} & \textbf{Worst result} & \textbf{Best result} \\
\              & min & avg & max & & \\
\hline
model 1 & 0.91 & 0.96 & 0.98 & 141048.jpg & 208078.jpg \\
model 2 & 0.91 & 0.96 & 0.98 & 141048.jpg & 208078.jpg \\
model 3 & 0.90 & 0.96 & 0.98 & 141048.jpg & 208078.jpg \\
ARCNN & 0.90 & 0.96 & 0.98 & 141048.jpg & 70011.jpg \\
input & 0.90 & 0.95 & 0.98 & 141048.jpg & 106005.jpg \\ 
\hline
\multicolumn{6}{c}{\textbf{Gain}} \\
\hline
model 1 & 0.00 & 0.01 & 0.02 & 156054.jpg & 250047.jpg \\
model 2 & -0.02 & 0.01 & 0.01 & 156054.jpg & 388006.jpg \\
model 3 & -0.02 & 0.01 & 0.01 & 156054.jpg & 102062.jpg \\
ARCNN & -0.01 & 0.01 & 0.01 & 156054.jpg & 48017.jpg \\
\hline
\end{tabular}
\label{tab:results_ssnr}
\end{center}
\end{table}
Graphs from Fig. \ref{fig:result_psnr} and Fig. \ref{fig:result_ssim}  show how much the results depend on the image itself and its specifics. It can be assumed that the algorithms improve the results of these measures but only slightly.
Some of the images after compression contain a significant distortion corresponding to the original image, which can only be partially corrected. The large diversity of images shows the irregular distribution of the PSNR and SSIM measure results for individual images. The SSIM results for all models are similar, which shows that they remove block artifacts similarly. The PSNR results, on the other hand, are divergent and show how models deal with noise that is an integral part of the original image. For Model 1, the predefined DCT layer weights can affect the least blur of small details of the corrected image.


\subsection{Problem of Lost Pattern}

To illustrate a high impact of the noise to the PSNR and SSIM values, we conducted an additional experiment. In this experiment, we compared the reconstructed image with the unchanged original image and with the slightly blurred one.  We apply a bilateral blur in a small range that will not change the sharp differences between the pixels but only equalize the slight background and texture noise.

\begin{figure*}
\centering
\includegraphics[width=\textwidth]{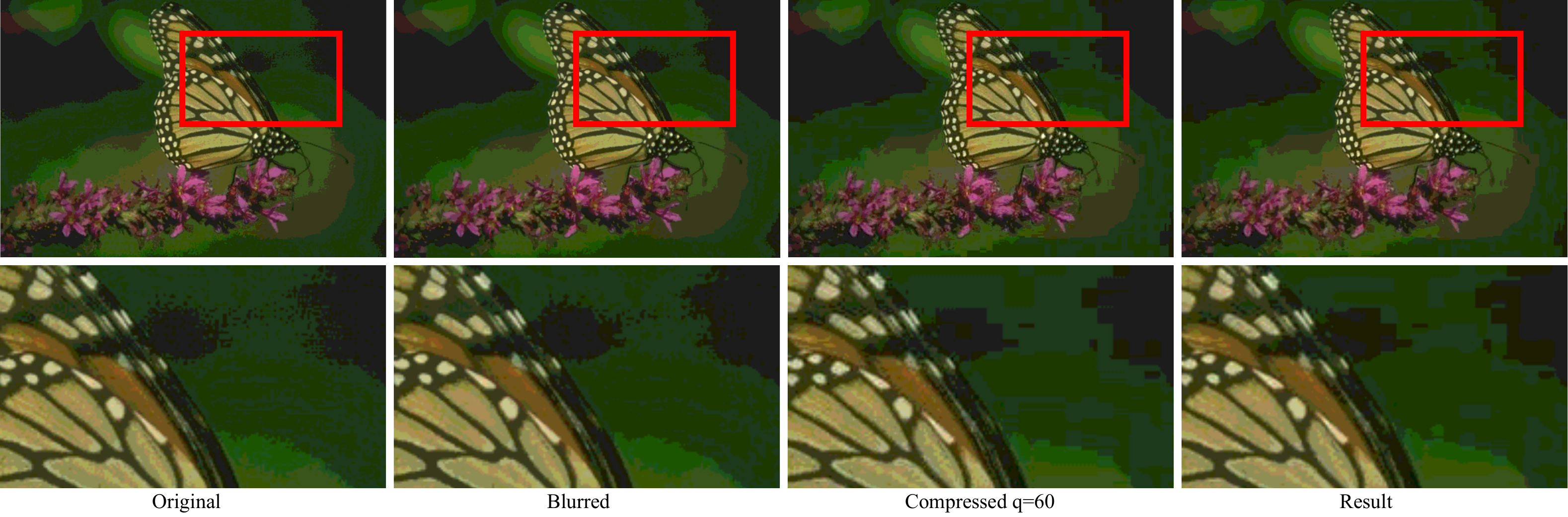}
\caption{The problem of natural noise losing after compression that influences the efficiency measures. The effect is visible after reducing the color space to 20 colors.}
\label{fig:grain_visualization}
\end{figure*}
\begin{figure}
\centering
\includegraphics[width=4.3cm]{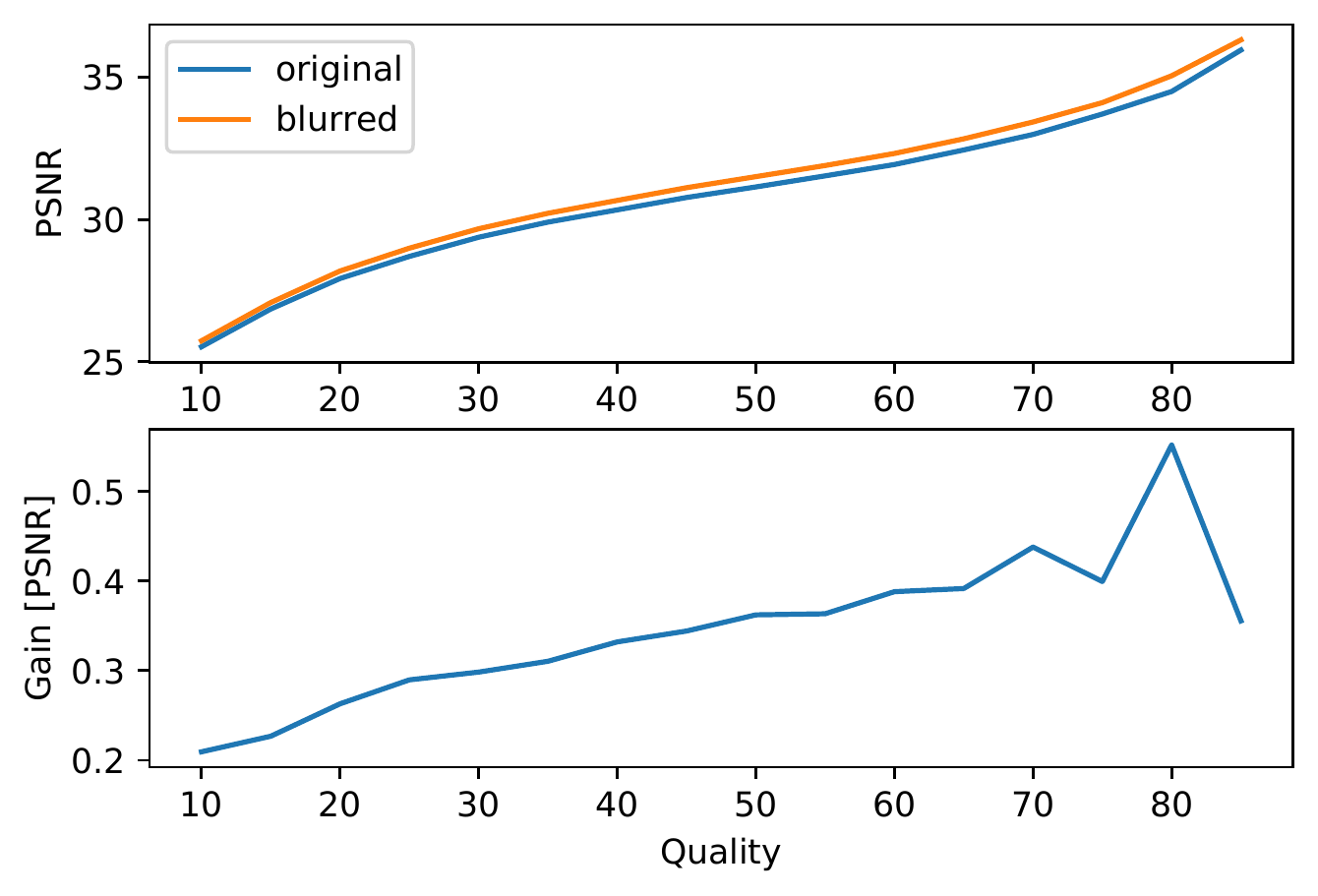}
\includegraphics[width=4.3cm]{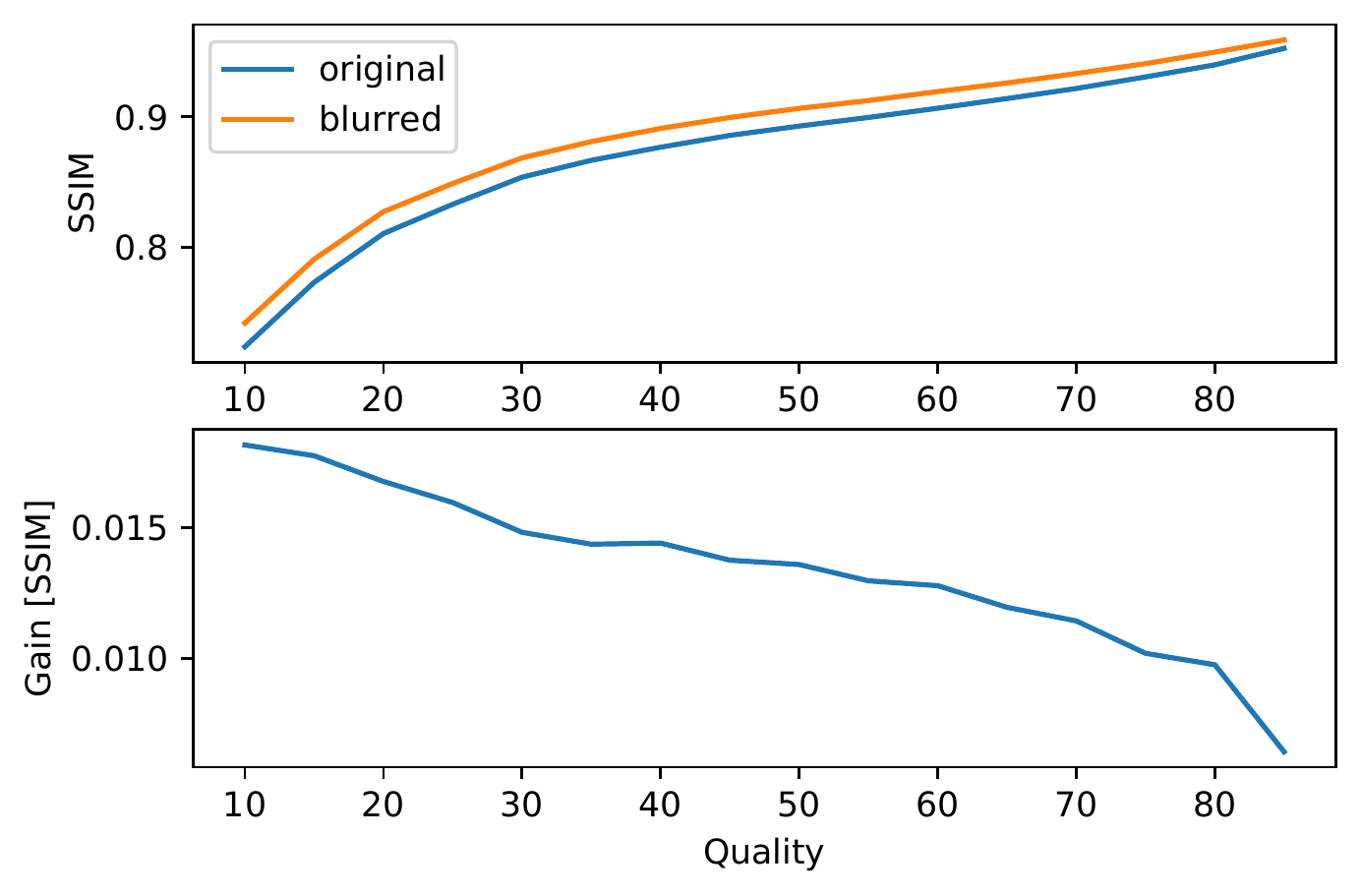}
\caption{Comparison of PSNR values between the compressed and the original image and its blurred version depending on the degree of the compression quality.}
\label{fig:encode_blure_psnr}
\end{figure}
\begin{figure}
\centering
\includegraphics[width=8cm]{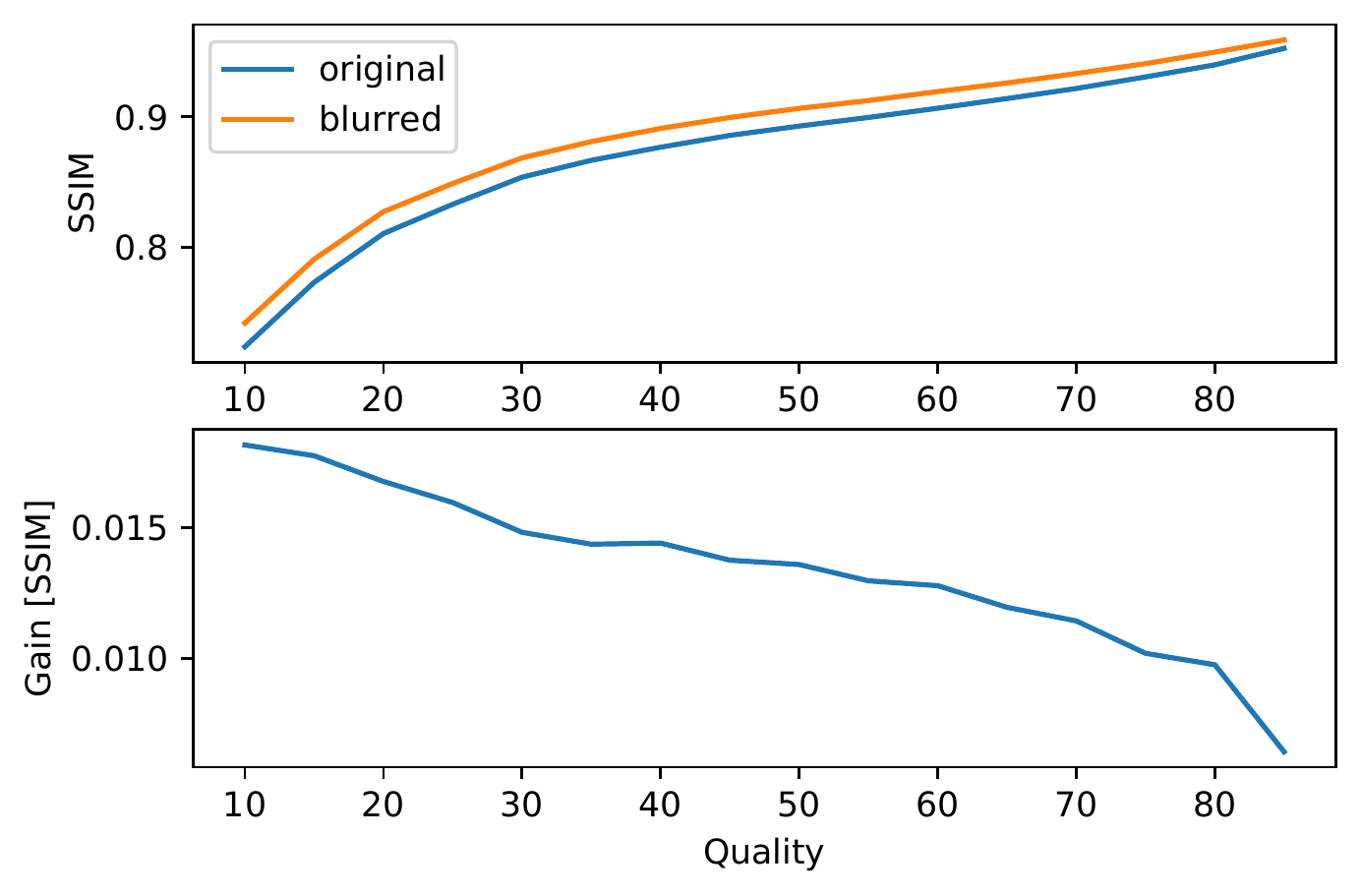}
\caption{Comparison of SSIM values between the compressed and the original image and its blurred version depending on the degree of the compression quality.}
\label{fig:encode_blure_ssim}
\end{figure}

\begin{figure}
\centering
\includegraphics[width=\columnwidth]{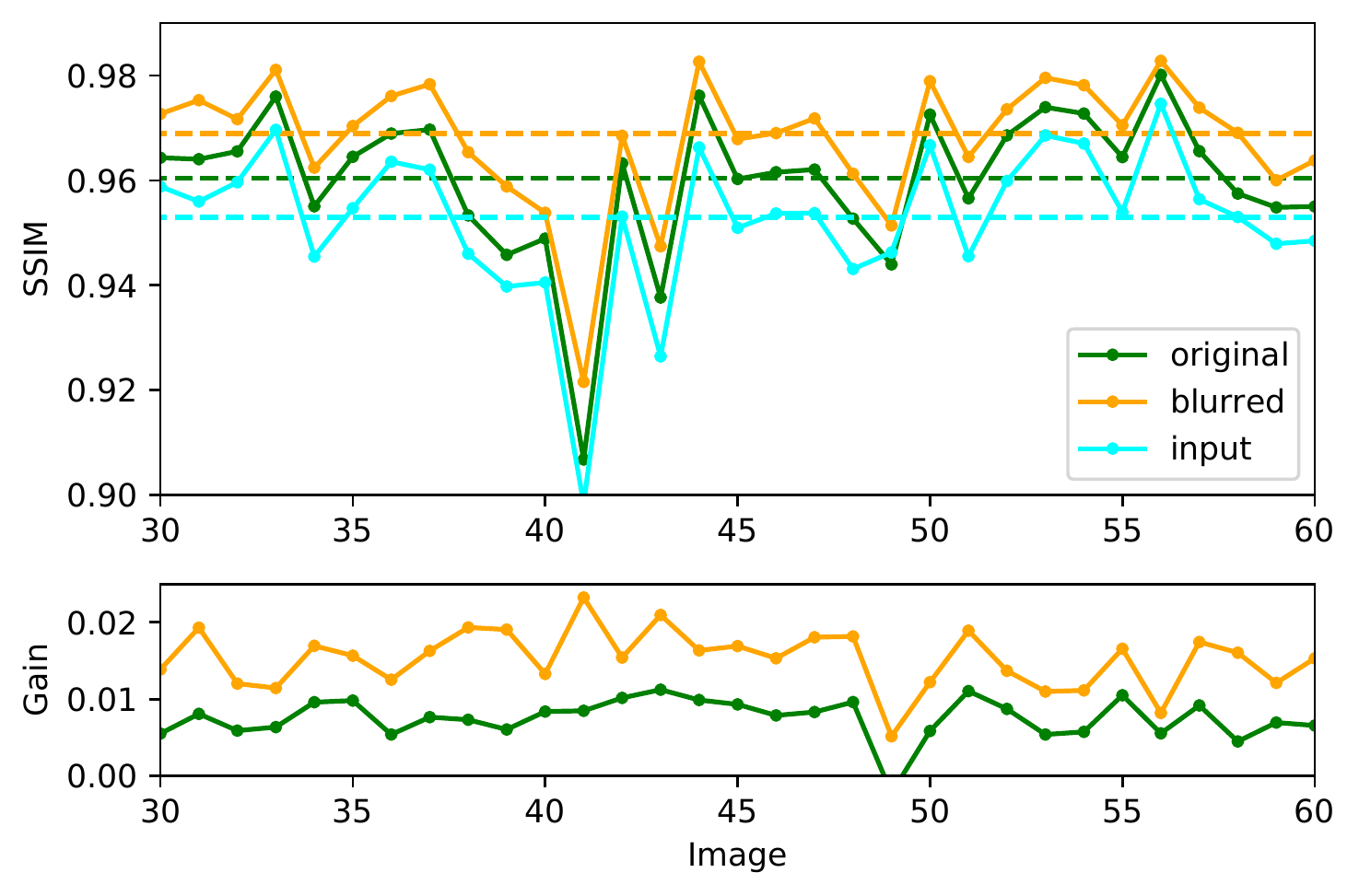}
\caption{Improvement of SSIM accuracy of artifact reduction measured for blurred images.}
\label{fig:predict_blurre_ssim}
\end{figure}

Fig. \ref{fig:encode_blure_psnr} and \ref{fig:encode_blure_ssim} show the differences in the PSNR and SSIM measure values between the original and blurred images. The increase in the PSNR gain is greater with a higher quality of compression, which means that the first thing that is lost is the slight noise.
Contrariwise with the SIMM measure, that blur improves the results for larger areas that are most degraded only at higher compression rates when blocks even on uniform areas become visible.
Noise is an element of the image that is irretrievably lost in compression. Fig. \ref{fig:grain_visualization} shows an example of such an action. After reducing the image colors to 10, the noise normally invisible begins to be noticed; however, immediately after compression, it is completely erased, and during reconstruction, it cannot be restored.  
\begin{figure}
\centering
\includegraphics[width=\columnwidth]{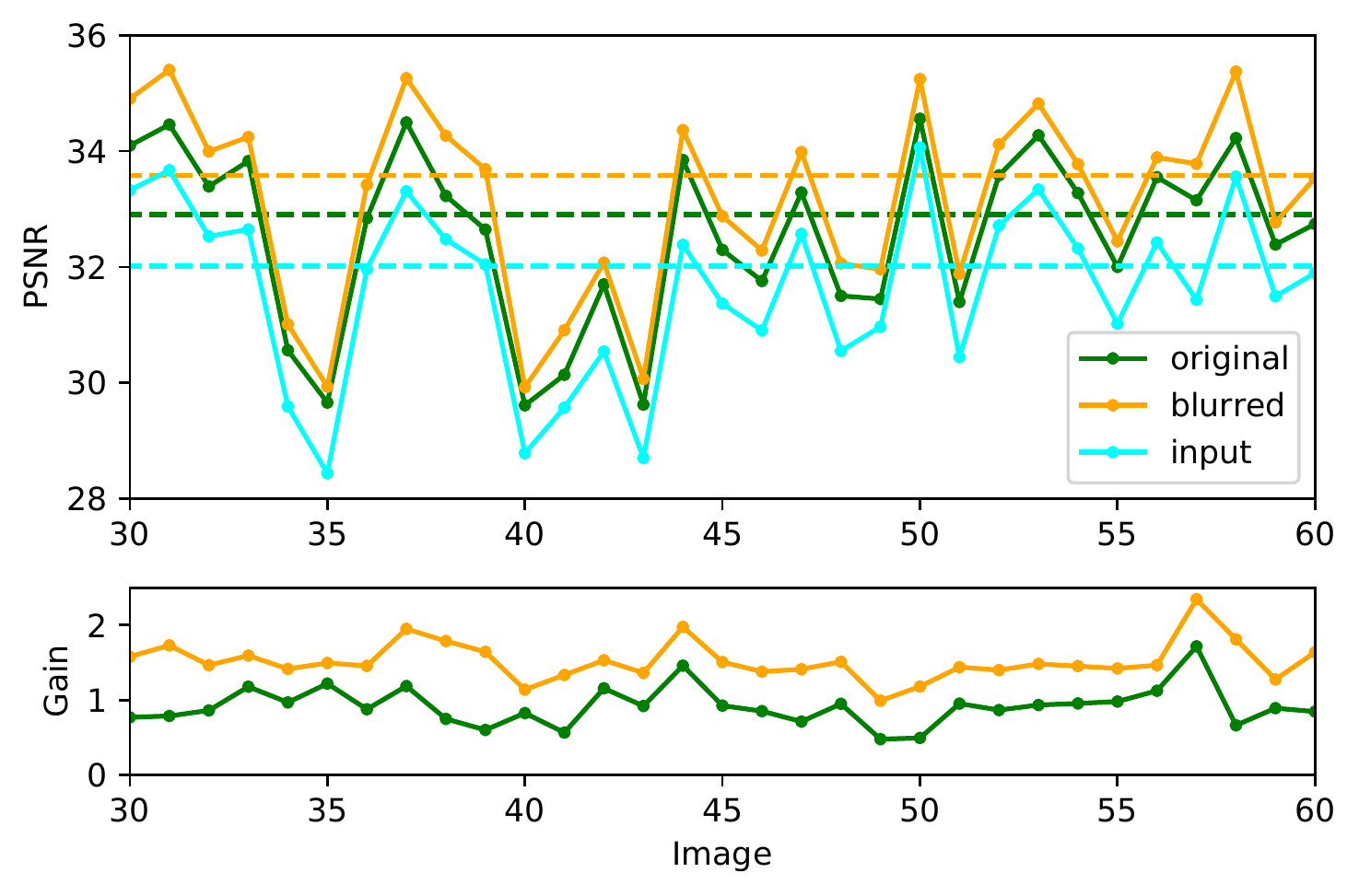}
\caption{Improvement of PSNR accuracy of artifact reduction measured for blurred images.}
\label{fig:predict_blurre_psnr}
\end{figure}
\begin{table}
\caption{Summary of the PSNR results after artifact removing in the case of comparison result with original image after blurring.}
\begin{center}
\begin{tabular}{ccccccc}
\hline 
\textbf{Model} & \multicolumn{3}{c}{\textbf{PSNR}} & \textbf{Worst result} & \textbf{Best result} \\
\              & min & avg & max & & \\
\hline
orginal & 28.02 & 32.90 & 39.09 & 346016.jpg & 70011.jpg \\
blurred & 28.42 & 33.58 & 40.03 & 346016.jpg & 70011.jpg \\
input & 27.43 & 32.01 & 37.79 & 346016.jpg & 43051.jpg \\
\hline
\multicolumn{6}{c}{\textbf{Gain}} \\
\hline
orginal & -0.19 & 0.89 & 2.14 & 196027.jpg & 48017.jpg \\
blurred & 0.56 & 1.57 & 2.66 & 196027.jpg & 48017.jpg \\
\hline
\end{tabular}
\label{tab:results_psnr_blurred}
\end{center}
\end{table}
\begin{table}[htbp]
\caption{Summary of SSIM results after artifacts removing in the case of comparison result with original image after bluring.}
\begin{center}
\begin{tabular}{ccccccc}
\hline 
\textbf{Model} & \multicolumn{3}{c}{\textbf{SSIM}} & \textbf{Worst result} & \textbf{Best result} \\
\              & min & avg & max & & \\
\hline
orginal & 0.91 & 0.96 & 0.98 & 141048.jpg & 208078.jpg \\
blurred & 0.92 & 0.97 & 0.99 & 346016.jpg & 106005.jpg \\
input & 0.90 & 0.95 & 0.98 & 141048.jpg & 106005.jpg \\
\hline
\multicolumn{6}{c}{\textbf{Gain}} \\
\hline
orginal & 0.00 & 0.01 & 0.02 & 156054.jpg & 250047.jpg \\
blurred & 0.01 & 0.02 & 0.03 & 156054.jpg & 71076.jpg \\
\hline
\end{tabular}
\label{tab:results_ssim_blurred}
\end{center}
\end{table}
Tables \ref{tab:results_psnr_blurred} and \ref{tab:results_ssim_blurred} show that the applied delicate blur improves image reconstruction results almost twice. Fig. \ref{fig:predict_blurre_psnr} and \ref{fig:predict_blurre_ssim} show how the reconstruction results change for individual images after blurring. The results mainly improve proportionally, only followed by slight deviations. The original and blurred image looks almost identical, while the increase in the value of measures is significant.

The noise problem is irreversible because its location has been lost, and therefore even an attempt to generate it ideally, e.g., by the GAN networks, would not give better results measured by the PSNR measure. The generated image would resemble the original, but the PSNR value will worsen because the noise between images will not converge and, in result, increase the variance between pixel values.
\section{Conclusion}
\label{sec:conclusions}
We presented new, fully convolutional networks for JPEG-compressed image improvement. In summary, the presented models obtain very similar results in removing JPEG compression artifacts. Experiments have shown that the use of predefined weights for the convolutional layer allowed for obtaining slightly better results compared to other networks with trained layers. This shows that in some cases, the so-called handmade filter can be more universal than learned one, especially when the problem is well known, like JPEG image compression.
In the case of artifacts removing in low-resolution images, the PSNR and SSIM measures become problematic and do not fully reflect the degree of elimination of artifacts. In the experiment, it turns out that the quality measures used commonly in the literature are very susceptible to noise, which does not play a significant role in image improvement.
The problem of noise is also different; an attempt to generate it artificially allows for better final image reception by the human, but the measure used shows worse results than if the noise was omitted.
%
\printbibliography
\end{document}